\documentclass[twocolumn,floatfix,showpacs]{revtex4}%
\usepackage{graphicx}%
\usepackage{amsmath}%
\usepackage{amsfonts}%
\usepackage{amssymb}
\usepackage{bm}
\usepackage{color}

\def\s{{\sigma}}
\def\e{{\epsilon}}
\def\k{{ {\bm k} }}
\def\p{{ {\bm p} }}
\def\q{{ {\bm q} }}

\def\0{{ {\bm 0} }}
\def\w{{\omega}}
\def\a{{\alpha}}
\def\b{{\beta}}

\allowdisplaybreaks[4]

\begin{document}
\title{
Superconductivity without hole-pocket in electron-doped FeSe: \\
Analysis beyond the Migdal-Eliashberg formalism
}
\author{
Youichi Yamakawa, 
and Hiroshi Kontani$^*$
}


\date{\today }

\begin{abstract}
High-$T_{\rm c}$ pairing mechanism absent of hole-pockets
in heavily electron-doped FeSe
is one of the key unsolved problems in Fe-based superconductors.
Here, this problem is attacked by focusing on the higher-order many-body effects
neglected in conventional Migdal-Eliashberg formalism.
We uncover two significant many-body effects 
for high-$T_{\rm c}$ superconductivity:
(i) Due to the ``vertex correction'',
the dressed multiorbital Coulomb interaction 
acquires prominent orbital dependence for low-energy electrons.
The dressed Coulomb interaction not only induces the orbital fluctuations,
but also magnifies the electron-boson coupling constant.
Therefore, moderate orbital fluctuations give strong 
attractive pairing interaction.
(ii) The ``multi-fluctuation-exchange pairing process''
causes large inter-pocket attractive force, which is 
as important as usual single-fluctuation-exchange process.
Due to these two significant effects
dropped in the Migdal-Eliashberg formalism,
the anisotropic $s_{++}$-wave state 
in heavily electron-doped FeSe is satisfactorily explained.
The proposed ``inter-electron-pocket pairing mechanism''
will enlarge $T_{\rm c}$ in other Fe-based superconductors.

\end{abstract}

\address{
Department of Physics, Nagoya University,
Furo-cho, Nagoya 464-8602, Japan. 
}
 

\sloppy

\maketitle

\section{Introduction}

The pairing mechanism in Fe-based superconductors
is a significant unsolved issue 
in condensed matter physics.
Strong spin and/or orbital fluctuations 
are expected to mediate the pairing interaction
 \cite{Kuroki-SC,Mazin,Hirschfeld-SC,Chubukov-SC,Kontani-RPA,Yin-SC,Hosono-Rev}.
To achieve a convincing answer on this problem, 
normal state electronic states should be clearly understood,
and therefore the electronic nematic state has been studied very actively.
The nematic state is the spontaneous rotational symmetry breaking 
driven by the electron correlation 
that triggers the small lattice distortion 
at $T=T_{\rm str}$.
Large orbital polarization ($E_{yz}-E_{xz}\sim60$meV)
is observed by the angle-resolved-photoemission spectroscopy (ARPES)
\cite{Yi-ARPES-Ba122, Zhang-ARPES-Na111}.
To explain the nematic phase in Fe-based superconductors,
various theoretical possibilities have been proposed so far.
Both the spin-nematic scenario
\cite{Kivelson,Fernandes} 
and the orbital/charge order scenario
\cite{Kruger-OO,Lee-OO,Onari-SCVC,Kontani-Raman} 
have been discussed very actively.

In the latter scenario,
it is difficult to derive the orbital/charge order 
based on the conventional mean-field-level theories,
like the random-phase-approximation (RPA).
This fact means that the higher-order electronic correlations,
called the vertex corrections (VCs),
should play essential roles on the nematic transition
\cite{Onari-SCVC}.
The Fermi liquid theory tells that
the dressed Coulomb interaction due to the VCs can
acquire nontrivial spin- and orbital-dependences for low-energy electrons
\cite{Schrieffer-UVC, Stamp-UVC}.
In Fe-based superconductors,
the orbital-dependent dressed Coulomb interaction
has been discussed in Refs.
\cite{Bascones-MF, Capone-Mott, Capone-Hund, Yu-Mott}.

In Ref. \cite{Onari-SCVC}, the authors found the 
significant role of the Aslamazov-Larkin (AL) type VC
for the bare Coulomb interaction ${\hat U}^0$,
which we call the $U$-VC in this paper.
Since the AL-type $U$-VC
enlarges the ``inter-orbital repulsive interaction''
under moderate spin fluctuations
\cite{Onari-SCVC,Yamakawa-FeSe,Onari-FeSe},
the orbital order is driven by the electron correlation 
in Fe-based superconductors.
In Refs. \cite{Yamakawa-CDW,Tsuchiizu2016-cuprate},
this mechanism has been applied to explain the
nematic charge-density-wave in cuprate superconductors,
whcih has been a significant open problem
in strongly-correlated electron systems
\cite{Sachdev-CDW, Wang-CDW}.

Considering the importance of the $U$-VC in the normal state
\cite{Onari-SCVC,Kontani-Raman},
the same $U$-VC should have strong impact on the superconductivity.
In conventional spin/orbital-fluctuation pairing theories,
the coupling constant between the electron and the fluctuations
(= bosons)
is simply given by the bare interaction ${\hat U}^0$ \cite{Kuroki-SC},
which we call the ``Midgal approximation'' in this manuscript.
However, the fluctuation-mediated pairing interaction is modified by the 
$U$-VCs for the electron-boson coupling constant,
which is expectet to be important in Fe-based superconductors.
Therefore,
we have to formulate the ``gap equation beyond the 
Migdal-Eliashberg (ME) theory'' 
by considering the $U$-VC seriously.
This is the main topic of the present study.

Among the Fe-based superconductors,
FeSe families attract considerable attention because of 
its high potential for realizing high-$T_{\rm c}$ superconducting state.
To explain the nematic state without magnetization in FeSe, 
the spin-nematic 
\cite{Valenti-FeSe,Wang-nematic,Yu-nematic} 
and the orbital-order
\cite{Yamakawa-FeSe,Onari-FeSe,Chubukov-nemaic2,Jiang-FeSe,Fanfarillo-FeSe} 
mechanisms have been discussed.
\cite{Baek,Ishida-NMR}
whereas strong ``nematic fluctuations'' 
are observed by the shear modulus and $B_{1g}$ electronic Raman study
\cite{Ishida-NMR,Raman-Gallais,Shibauchi-nematic}.
Below $T_{\rm str}$, large orbital polarization appears 
with the unconventional sign-reversal in $\k$-space
\cite{FeSe-ARPES6,Onari-FeSe,Jiang-FeSe,Fanfarillo-FeSe}.
These characteristic non-magnetic nematic states above and below $T_{\rm str}$
are quantitatively explained by considering the $U$-VC,
using the self-consistent VC (SC-VC) theory
\cite{Yamakawa-FeSe,Onari-FeSe}.

In FeSe, high-$T_{\rm c}$ state emerges
by introducing $10\sim15$\% electron carrier,
as observed in ultrathin FeSe on SrTiO$_3$ ($T_{\rm c}=40\sim100$K)
\cite{eFeSe1,eFeSe5,eFeSe6,Feng-eFeSe-swave,Shen-eFeSe},
K-coated FeSe ($T_{\rm c}\lesssim50$K) \cite{Takahashi-eFeSe,Feng-eFeSe},
and intercalated superconductor (Li$_{0.8}$Fe$_{0.2}$)OHFeSe 
($T_{\rm c}\lesssim40$K)
\cite{Feng-eFeSe-swave2,HHWen-eFeSe}.
The fully-gapped $s$-wave state has been confirmed experimentally 
\cite{Feng-eFeSe-swave,Shen-eFeSe,HHWen-eFeSe}, and the 
sign-preserving $s_{++}$-wave state is reported in Refs.
\cite{Feng-eFeSe-swave,Feng-eFeSe-swave2}.
In these high-$T_{\rm c}$ compounds,
the top of the hole-pocket completely sinks below the 
Fermi level ($\sim -0.1$eV).
Although strong interfacial electron-phonon interaction
may increase $T_{\rm c}$ in FeSe film on SrTiO$_3$
\cite{Shen-replica,Millis,DHLee,Choi,Jhonston},
its importance is not clear for other electron-doped FeSe. 
Thus, it is a significant unsolved problem to explain
the high-$T_{\rm c}$ $s$-wave state without hole-pockets
based on the repulsive Coulomb interaction.

In this paper,
we study the high-$T_{\rm c}$ pairing mechanism for 
heavily  electron-doped ($e$-doped) FeSe.
Even in the absence of the hole-pocket, 
we find that moderate spin and orbital fluctuations develop,
due to the orbital-spin interplay through the $U$-VC.
Then, moderate orbital fluctuations give rise to 
strong attractive pairing interaction 
since the electron-boson coupling constant is dressed and magnified
by the $U$-VC.
In addition, the AL-type multi-fluctuation-exchange pairing process 
causes large inter-pocket attractive force, which is not simply
proportional to the  single-fluctuation-exchange process.
Because of these beyond-ME pairing mechanisms 
uncovered by the present study,
the fully-gapped $s_{++}$-wave state is naturally obtained.
The obtained anisotropic gap structure is 
consistent with experimental results
\cite{Shen-eFeSe,Feng-eFeSe-swave,Feng-eFeSe-swave2,HHWen-eFeSe}.

The $s_{++}$-wave state in heavily $e$-doped compounds 
due to the charge-channel fluctuations
was discussed based on phenomenological approaches
\cite{Saito-KFe2As2,Fernandes-eFeSe}. 
Here, we analyze the realistic Hubbard model 
using the advanced microscopic theory,
and uncover the beyond-ME pairing mechanism
responsible for the strong attractive pairing interaction.

The nematic orbital order is driven by the combination 
of the VC due to the electron correlation
and the electron-phonon interaction.
In Fe-based superconductors, the  electron correlation effect is the 
main driving force of the nematic transition,
as theoretically explained in Ref.
\cite{Kontani-Raman}.

\section{Vertex corrections in Multiorbital Systems
based on the Fermi liquid theory}

\begin{figure}[!htb]
\includegraphics[width=.9\linewidth]{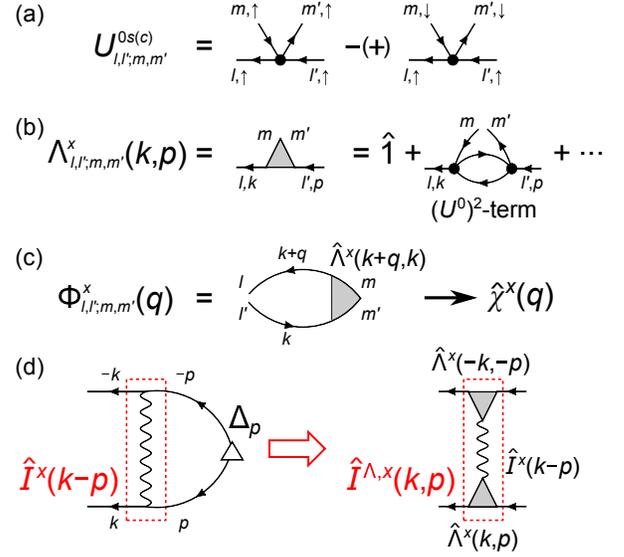}
\caption{
(color online)
(a) The Hubbard interaction for the spin- (charge-) channel
${\hat U}^{0s(c)}$, denoted as the black circle.
(b) The three-point VC for the coupling constant
between the electron and the fluctuations ($U$-VC), 
$\Lambda_{l,l';m,m'}^x(k,p)$ ($x=s,c$).
(${\hat \Lambda}^x={\hat 1}$ for $U\rightarrow 0$.)
The ($U^0$)-linear terms are dropped since they are included in the RPA diagrams.
The dressed coupling constant modified by the $U$-VC is
${\hat U}^{x}(k,p)={\hat \Lambda}^x(k,p){\hat U}^{0x}$.
(c) Beyond the RPA:
${\hat \Phi}^x(q)$ is the irreducible susceptibility with the VC.
(d) Single-fluctuation-exchange pairing interaction
beyond the ME approximation.
The $U$-VC is given by Eq. (\ref{eqn:beyondME}).
}
\label{fig:fig1}
\end{figure}

We first introduce the $U$-VC,
which expresses the three-point vertex
connecting between the electron and the fluctuations.
We show that the $U$-VC
strongly modifies the essential electronic properties
in strongly correlated electron systems.
In multiorbital systems,
the Coulomb interaction is represented by
the intra-orbital term $U$, 
the inter-orbital term $U'$,
and the exchange or Hund's coupling term $J$.
Its matrix expression, ${\hat U}^{0x}$,
is given in Fig. \ref{fig:fig1} (a)
\cite{Kuroki-SC,Onari-SCVC},
where $x=c,s$ represents the charge- or spin-channel interaction.
Its expression for Fe-based superconductors 
is given in the next section.
(Note that $U^{0s}=-U^{0c}=U$ in the single-orbital case.)

Due to the many-body effect,
the model interaction is changed to 
energy- and momentum-dependent dressed interaction ${\hat U}^{x}(k,p)$,
where $k=(\e_n=(2n+1)\pi T, \k)$ and $p=(\e_m, \p)$.
The lowest few terms for the three-point vertex ($U$-VC)
are shown in Fig. \ref{fig:fig1} (b).
The solid lines are the electron Green functions.
The dressed interaction is expressed as
${\hat \Lambda}^x(k,p){\hat U}^{0x}$, and 
we call ${\hat \Lambda}^x(k,p)$ the $x$-channel $U$-VC.
Note that the $U$-VC is irreducible with respect to ${\hat U}^{0x}$,
and $\Lambda^x_{l,l';m,m'}=\delta_{l,m}\delta_{l',m'} \equiv {\hat 1}$
in the Migdal approximation.

The significance of the $U$-VC in Fe-based superconductors
was discovered in Ref. \cite{Onari-SCVC}:
Strong orbital fluctuations arise from the $U$-VC that is
included in the irreducible charge susceptibility 
\begin{eqnarray}
{\hat \Phi}_{l,l';m,m'}^x(q)&=&-T\sum_{k,l_1,l_2}{G}_{l,l_1}(k+q){G}_{l_2,l'}(k)
 \nonumber \\
& &\times{\Lambda}_{l_1,l_2;m,m'}^x(k+q,k) ,
\label{eqn:Phi}
\end{eqnarray}
which is illustrated in Fig. \ref{fig:fig1} (c).
Here, $q=(\w_l=2\pi l T, \q)$.
Then, the susceptibility
\begin{eqnarray}
{\hat \chi}^x(q)
={\hat \Phi}^x(q)[{\hat 1}-{\hat U}^{0x}{\hat \Phi}^x(q)]^{-1}
 \ \ \ (x=s,c) ,
\label{eqn:chi}
\end{eqnarray}
is strongly influenced by the $U$-VC included in ${\hat \Phi}^x(q)$.
The magnetic (orbital) order occurs when the 
spin (charge) Stoner factor $\a_{S}$ ($\a_C$), which is given as the 
maximum eigenvalue of ${\hat U}^{s(c)}{\hat \Phi}^{s(c)}(q)$, reaches unity.

The $U$-VC should also be significant for the superconductivity:
Conventionally,
the pairing interaction due to the single-fluctuation-exchange
(=Maki-Thompson process) is studied, and the $U$-VC 
for the electron-boson coupling is dropped (=Migdal approximation).
Then, the interaction due to charge or spin susceptibility
$\chi^{x}$ is
\begin{eqnarray}
{\hat I}^{x}(k-p)
= {\hat U}^{0x}+{\hat U}^{0x}{\hat \chi}^{x}(k-p){\hat U}^{0x}
\label{eqn:ME} ,
\end{eqnarray}
where $x=c$ (charge) or $x=s$ (spin).
However,
the bare electron-boson coupling 
should be dressed by the $U$-VC as shown in Fig. \ref{fig:fig1} (d),
which is required by the microscopic Fermi liquid theory
\cite{Tazai}.
The pairing interaction with the $U$-VC 
(=beyond Migdal approximation) is given as
\begin{eqnarray}
{\hat I}^{\Lambda,x}(k,p)= {\hat \Lambda}^x(k,p){\hat I}^{x}(k-p)
{\hat{ \bar \Lambda}}^x(-k,-p),
\label{eqn:beyondME}
\end{eqnarray}
where ${\bar \Lambda}^x_{l,l';m,m'}(k,p)\equiv \Lambda^x_{m',m;l',l}(k,p)$.
Its expression is illustrated in Fig. \ref{fig:fig1} (d).
We stress that ${\hat \Lambda}^x$ and ${\hat \Phi}^x$
are exactly related by the one-to-one relationship
${\hat \Phi}^x(q)=-T\sum_k{\hat G}(k+q){\hat G}(k){\hat \Lambda}^x(k+q,k)$.


\section{Model Hamiltonian for 15\% e-doped FeSe}

In this paper, we analyzed the 
eight-orbital Hubbard model for heavily $e$-doped FeSe model:
%
\begin{eqnarray}
H= H_0 + r H_U .
\label{eqn:Ham}
\end{eqnarray}
Here, $H_0$ is the kinetic term, which we will introduce below,
and $H_U$ is the first-principles multiorbital interaction for FeSe
\cite{Arita}.
The factor $r$ is the reduction factor for the interaction term
\cite{Yamakawa-FeSe}.

The kinetic term $H_0$ is given as
\begin{eqnarray}
H_0 = \sum_{\k,ll'\s} c^\dagger_{\k,l\s}H^{0,{\rm b}}_{\k,l,l'}c_{\k,l'\s}
+ \Delta H^0
\label{eqn:H0}
\end{eqnarray}
where 
$H^{0,{\rm b}}_{\k,l,l'}$ is the tight-binding model
for the bulk FeSe introduced in Ref. \cite{Yamakawa-FeSe}.
Here, $l$ is the orbital index, and we denote 
$d_{z^2},d_{xz},d_{yz},d_{xy},d_{x^2-y^2}$ orbitals as 1, 2, 3, 4, 5,
and $p_x,p_y,p_z$ orbitals as 6, 7, 8.
In bulk FeSe, the relation $E_{xy}<E_{yz}(<\mu)$ holds at X point.
In $e$-doped FeSe, however, the opposite relation $E_{xy}>E_{yz}$ holds,
and the Fermi velocity is much smaller
\cite{Shen-replica,JJSeo}.
We introduce $\Delta H^0$
in order to reproduce the experimental bandstructure and FSs in $e$-doped FeSe,
we shift the $E_{xz}$ [$E_{xy}$] level at 
$\k=((0,0)$, $(\pi/2,0)$, $(\pi,0)$, $(0,\pi/2)$, $(\pi/2,\pi/2)$, 
$(\pi,\pi/2)$, $(0,\pi)$, $(\pi/2,\pi)$, $(\pi,\pi))$ by
$(+0.2$, $0$, $0$, $0$, $0$, $0$, $+0.2$, $0$, $0$) 
[$(+0.25$, $-0.1$, $+0.4$, $-0.1$, $0$, $0$, $+0.4$, $0$, $0$)] in unit eV,
by introducing the additional intra-orbital hopping integrals for $l=2\sim4$.

\begin{figure}[!htb]
\includegraphics[width=.99\linewidth]{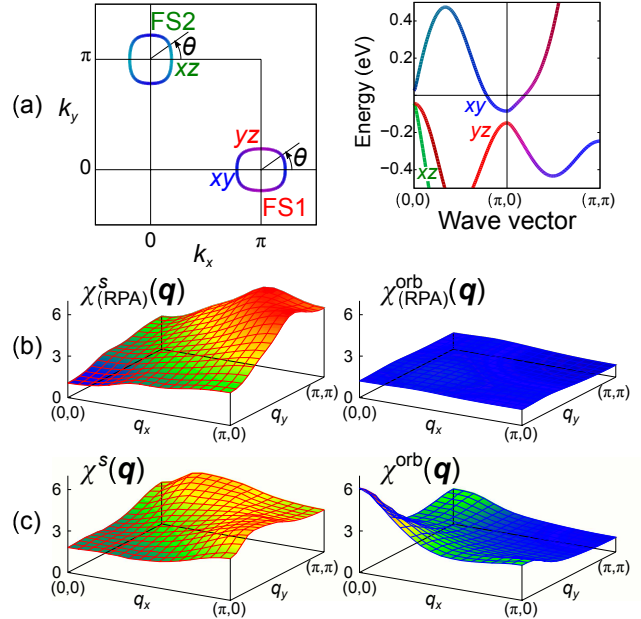}
\caption{
(color online)
(a) FSs and bandstructure of the 1Fe-UC FeSe model
with 15\% $e$-doping.
Green, red, and blue lines correspond to $xz$, $yz$, and $xy$ orbitals, 
respectively.
In each FS, the $xy$-orbital has large weight
near the $k_x$- or $k_y$-axis.
(b) The spin and orbital susceptibilities obtained by the RPA.
The latter is very small in the RPA.
(c) The spin and orbital susceptibilities 
obtained by the SC-VC theory.
The $U$-VCs for the irreducible susceptibilities ${\hat \Phi}^{s,c}(q)$
are calculated self-consistently.
Moderate ferro-orbital fluctuations are 
induced by the charge-channel $U$-VC,
consistently with the experimental phase diagram of 
$e$-doped FeSe.
}
\label{fig:fig2}
\end{figure}

Figure \ref{fig:fig2} (a) shows the bandstructure and FSs,
which are unfolded in the present 1-Fe unit cell (1Fe-UC) model.
In this paper, we introduce the orbital-dependent renormalization factors
$z_4=1/1.6$ and $z_l=1$ ($l\ne4$)
 \cite{Yamakawa-FeSe}:
Consistently, the relation $z_{2,3}/z_4 \sim 1.3$ is given by the 
dynamical-mean-field-theory in Ref. \cite{DMFT1}.
The band dispersion is given by the pole of the Green function
${\hat G}(\k,\e)= ({\hat Z}\cdot\e-{\hat H}_0(\k))^{-1}$,
where ${Z}_{l,m}=(1/z_l)\delta_{l,m}$ \cite{Yamakawa-FeSe,Onari-FeSe}.
Equivalently, it is given by the eigenvalues of
${\hat Z}^{-1/2}{\hat H}_0(\k){\hat Z}^{-1/2}$.

The Coulomb interaction term for $d$-electrons $H_U$ is expressed as
\begin{equation}
H_U = -\frac12 \sum_{i,ll'mm'}\sum_{\s\rho}
U^0_{l\s,l'\s;m\rho,m'\rho} c^\dagger_{i,l\s}c_{i,l'\s}c^\dagger_{i,m'\rho}c_{i,m\rho},
\end{equation}
where 
\begin{eqnarray}
U^0_{l\s,l'\s';m\rho,m'\rho'}
&=&\frac12 {U}^{0c}_{l,l';m,m'}\delta_{\s,\s'}\delta_{\rho',\rho}
\nonumber \\
&&+\frac12 {U}^{0s}_{l,l';m,m'}{\bm \s}_{\s,\s'}\cdot {\bm \s}_{\rho',\rho}
\label{eqn:SU2} ,
\end{eqnarray}
where ${\bm \s}=(\s_x,\s_y,\s_z)$ is the Pauli matrix vector.
The relationship with respect to the spin indices in Eq. (\ref{eqn:SU2}) 
is rigorous even for the dressed four-point vertex function,  
by reflecting the $SU(2)$ symmetry in the absence of the SOI
\cite{Tazai}.
In addition, the relationship
${\hat U}^{0\mathrm{s}}_{l.l';m,m'}
-{\hat U}^{0\mathrm{c}}_{l.l';m,m'} 
=2{\hat U}^{0\mathrm{s}}_{l.m;l',m'}$ holds.

${\hat U}^{0s}$ and ${\hat U}^{0c}$ is the matrix expression
for the multiorbital interaction for the spin or charge channel
introduced in Refs. \cite{Kuroki-SC,Onari-SCVC,Yamakawa-FeSe}:
\begin{equation}
{U}^{0\mathrm{s}}_{l,l';m,m'} = \begin{cases}
U_{l,l}, & l=l'=m=m' \\
U_{l,l'}' , & l=m \neq l'=m' \\
J_{l,m}, & l=l' \neq m=m' \\
J_{l,l'}, & l=m' \neq l'=m \\
0 , & \mathrm{otherwise},
\end{cases}
\end{equation}
and
\begin{equation}
{U}^{0\mathrm{c}}_{l,l';m,m'} = \begin{cases}
-U_{l,l}, & l=l'=m=m' \\
U_{l,l'}'-2J_{l,l'} , & l=m \neq l'=m' \\
-2U_{l,m}' + J_{l,m} , & l=l' \neq m=m' \\
-J_{l,l'} , & l=m' \neq l'=m \\
0 . & \mathrm{otherwise}.
\end{cases}
\end{equation}
Here, $U_{l,l}$, $U_{l,l'}'$ and $J_{l,l'}$
are the first-principles Coulomb interaction terms for FeSe
obtained in Ref. \cite{Arita}.

\section{Orbital and spin susceptibilities}

From now on, we perform the numerical study for 
the spin- and orbital susceptibilities in the 15\% e-doped FeSe model
shown in Fig. \ref{fig:fig2} (a).
They are given by Eqs. (\ref{eqn:Phi}) and (\ref{eqn:chi}).
In the RPA, 
the $U$-VC is dropped (${\hat \Lambda}^x={\hat 1}$), so
${\hat \Phi}^x(q)$ is reduced to the bare susceptibility
$\chi_{l,l';m,m'}^0(q)= -T\sum_k G_{l,m}(k+q)G_{m',l'}(k)$.
In this case, the relation $\a_S>\a_C$ holds, and therefore
the non-magnetic orbital order failed to be explained.

In Fig. \ref{fig:fig2} (b), we show the 
the RPA results for the total spin susceptibility,
$\chi^{s}(\q)\equiv \sum_{l,m}^{1\sim5}\chi_{l,l;m,m}^{s}(\q)$,
and the orbital susceptibility for the operator $O\equiv n_{xz}-n_{yz}$,
$\chi^{\rm orb}(\q)\equiv \sum_{l,m}^{2,3}(-1)^{l+m}\chi_{l,l;m,m}^{c}(\q)$.
The model parameters are $r=0.26$ and $T=30$ meV.
The spin and charge Stoner factors are $(\a_S,\a_C)=(0.80,0.48)$.
The broad incommensurate peak in $\chi^{s}_{\rm (RPA)}(\q)$
originates from the multiple nesting vectors,
such as the inter- and intra-pocket nestings
in addition to the contribution from the hole-pockets
sink below the Fermi level.
$\chi^{\rm orb}_{\rm (RPA)}(\q)$ remains small in the RPA.

\begin{figure}[!htb]
\includegraphics[width=.9\linewidth]{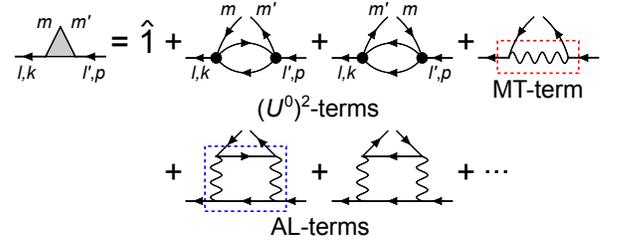}
\caption{
(color online)
Diagrammatic expression for the $U$-VC
analyzed in the present theoretical study.
Double counting terms are subtracted carefully.
The MT (AL) diagram inside the red (blue) dotted rectangle
corresponds to the MT (AL) diagram in the pairing interaction 
in Fig. \ref{fig:fig1} (d) (Fig. \ref{fig:fig4} (e)).
}
\label{fig:Lambda-x}
\end{figure}

By going beyond the RPA,
the opposite relation $\a_C>\a_S$ can be realized
when the $U$-VC for the charge channel becomes greater than ${\hat 1}$.
In the SC-VC theory, the diagrammatic expression for ${\hat \Lambda}^x$ 
is shown in Fig. \ref{fig:Lambda-x},
in which all the diagrams up to $O((\chi^{x})^2)$ 
are calculated self-consistently.
The analytic expressions for $U$-VC are explained in Appendix A.
Based on the SC-VC theory, we can explain the strong development of the 
orbital fluctuations in Fe-based superconductors
since ${\hat \Phi}^c(q)$ for the $d_{xz/yz}$-orbitals 
is enlarged by the charge-channel $U$-VC 
\cite{Onari-SCVC,Yamakawa-FeSe}.

The RPA results are qualitatively modified by the $U$-VC:
In Fig. \ref{fig:fig2} (c), 
we show $\chi^s(\q)$ and $\chi^{\rm orb}(\q)$ obtained by the SC-VC theory
for $r=0.35$ at $T=30$ meV, 
where the obtained Stoner factors are $(\a_S,\a_C)=(0.83,0.81)$.
Due to AL type $U$-VC in $\Phi^c$,
$\chi^{c}(\q)$ shows moderate peak at $\q\sim(0,0)$,
consistently with the experimental phase diagram of $e$-doped FeSe 
\cite{Feng-eFeSe}.
Also, the $\q$-dependence of $\chi^{s}(\q)$ is weakly
modified from the RPA result due to the $U$-VC in $\Phi^s$.
More detailed numerical results obtained by the SC-VC theory 
are explained in Appendix B.

\section{Gap equation and pairing interaction beyond the ME formalism}

Here, we study the following linearized gap equation
in the band-diagonal basis:
%
\begin{eqnarray}
\lambda\Delta_\a(k)=-T\sum_{p,\b}V_{\a,\b}^{\rm SC}(k,p)|G_\b(p)|^2\Delta_\b(p),
\label{eqn:gapeq}
\end{eqnarray}
where $\Delta_\a(k)$ is the gap function on the $\a$-band,
$\lambda$ is the eigenvalue, which is proportional to $T_{\rm c}$, and
$\lambda=1$ is satisfied at $T=T_c$.
$V_{\a,\b}^{\rm SC}$ is the pairing interaction in the band-diagonal basis.
For the singlet case, the pairing interaction with the $U$-VC is 
given by 
%
\begin{eqnarray}
{\hat V}^{\Lambda}(k,p)=\frac32 {\hat I}^{\Lambda,s}(k,p)
-\frac12 {\hat I}^{\Lambda,c}(k,p)-{\hat U}^{0s}
\label{eqn:V-UVC} ,
\end{eqnarray}
where $I^{\Lambda,x}$
is given in Eq. (\ref{eqn:beyondME}) 
using the SC-VC susceptibilities.
The ($U^0$)-linear term of Eq. (\ref{eqn:V-UVC})
is $\frac12[{\hat U}^{0s}-{\hat U}^{0c}]$.

\begin{figure}[!htb]
\includegraphics[width=.99\linewidth]{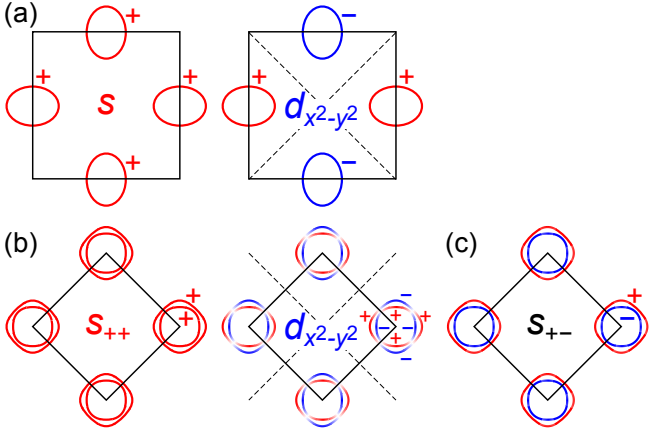}
\caption{
(color online)
(a) The schematic fully-gapped $s$-wave and $d$-wave states
in the 1Fe-UC Brillouin zone.
For former (latter) is realized when the inter-pocket pairing interaction 
is attractive (repulsive).
(b) The $s_{++}$-wave state and $d$-wave state in the 2Fe-UC Brillouin zone.
The inner FS and outer FS are formed due to the 
SOI-induced band hybridization.
In the $d$-wave state, the gap structure becomes nodal and 
$T_{\rm c}$ is suppressed due to the SOI-induced hybridization \cite{DHLee}.
(c) The $s_\pm$-wave state discussed in Ref. \cite{Chubukov-KSe}.
}
\label{fig:fig3}
\end{figure}

Before analyzing Eq. (\ref{eqn:gapeq}) numerically, 
we briefly discuss the possible gap states in heavily e-doped FeSe
\cite{DHLee,Fernandes-eFeSe}:
Since the hole-FSs are absent,
the only possible states are the $s$-wave and $d$-wave states,
shown in Fig. \ref{fig:fig3} (a).
The former (latter) state is realized 
when the inter-pocket pairing interaction is attractive (repulsive),
which may be mediated by the orbital (spin) fluctuations.
Next, we study the realistic 2Fe-UC model
to understand the effect of the spin-orbit interaction (SOI) 
$\lambda_{\rm SOI} {\bm l}\cdot{\bm s}$.
Due to the SOI-induced hybridization between FS1 and FS2,
the $d$-wave state possesses the nodal gap structure 
as schematically shown in Fig. \ref{fig:fig3} (b),
accompanied by drastic decrease in $T_{\rm c}$
\cite{DHLee}.
On the other hand, the $s$-wave state 
is essentially insensitive to the SOI-hybridization.
In addition, the $s_\pm$-wave state in Fig. \ref{fig:fig3} (c)
was discussed theoretically \cite{Chubukov-KSe}.


\begin{figure}[!htb]
\includegraphics[width=.99\linewidth]{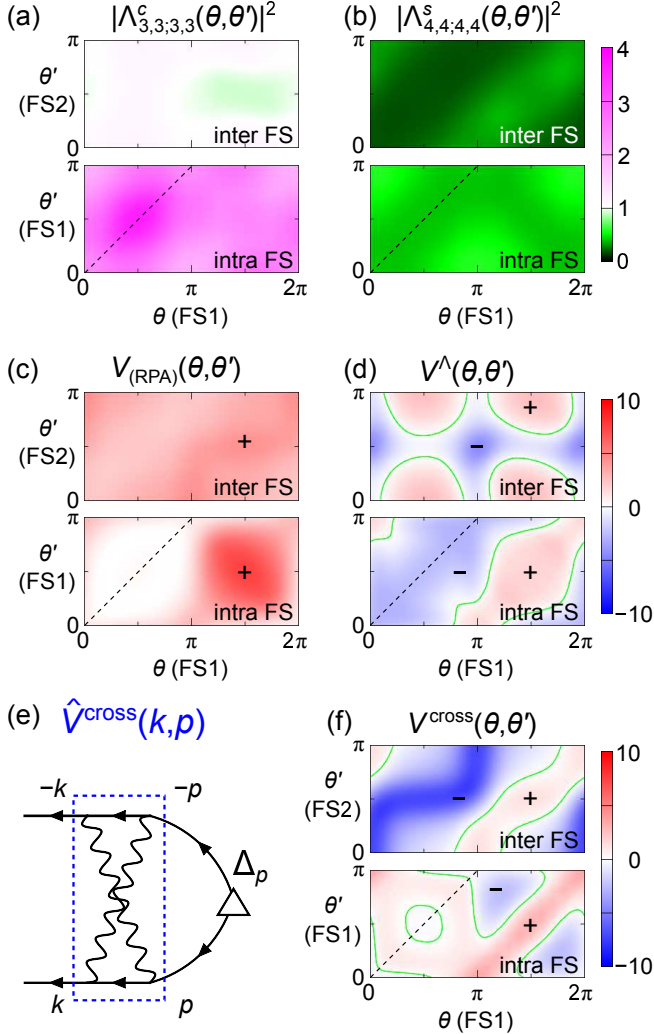}
\caption{
(color online)
(a) $|\Lambda^c_{3,3;3,3}(\theta,\theta')|^2 \ (\gg 1)$ and 
(b) $|\Lambda^s_{4,4;4,4}(\theta,\theta')|^2 \ (\ll 1)$ in $e$-doped FeSe model
at the lowest frequency ($\e_n=\e_n'=\pi T$),
where $\theta$ and $\theta'$ represent the Fermi momenta;
see Fig. \ref{fig:fig2} (a).
We show the pairing interactions (c) $V_{\rm (RPA)}$ and (d) $V^\Lambda$ on the FSs.
The summation for the lowest Matsubara frequencies
($\e_n=\pm\e_{n'}=\pi T$) is taken.
The attractive interaction in (d) originates from the orbital fluctuations.
(e) The crossing-fluctuation-exchange (=AL process) 
pairing interaction $V^{\rm cross}(k,p)$,
which represents the multi-fluctuation-exchange processes.
(f) $V^{\rm cross}$ on the FSs, in which
strong attractive inter-pocket interaction gives the $s$-wave state.
}
\label{fig:fig4}
\end{figure}

Here, we explain the significance of the $U$-VCs
for the pairing interaction in Eq. (\ref{eqn:beyondME}).
In Fig. \ref{fig:fig4}, we show the obtained
(a) $|\Lambda^c_{3,3;3,3}|^2$ and 
(b) $|\Lambda^s_{4,4;4,4}|^2$ on the FSs
in 15\% $e$-doped FeSe model,
in the case of $r=0.35$ [$(\a_S,\a_C)=(0.80,0.73)$] at $T=30$ meV.
Their analytic expressions are given in Appendix A.
In these figures,
$\theta$ and $\theta'$ represent the momenta on the FSs
$\k$ and $\p$ respectively.
It is found that $|\Lambda^c_{3,3;3,3}(\k,\p)|^2\lesssim4$ 
for the intra-pocket ($\k,\p\in$ FS1).
Then, $V^{\Lambda,c}$ gives strong attractive interaction
under moderate spin fluctuations.
Such large charge-channel $U$-VC originates from the AL-type VC,
consistently with the functional-renormalization-group 
analysis in Ref. \cite{Tazai} 
based on the functional-renormalization-group method.
In contrast, the opposite relation holds for the spin-channel $U$-VC;
$|\Lambda^s_{4,4;4,4}(\k,\p)|^2 \lesssim 0.35$
for the inter-pocket ($\k\in$ FS1, $\p\in$ FS2).
Thus, spin-fluctuation-mediated 
inter-pocket repulsion is reduced by the spin-channel $U$-VC
 \cite{Tazai}.

Next, we study the pairing interaction.
In the RPA without any $U$-VC, $V_{\rm (RPA)}(\k,\p)$,
both intra- and inter-pocket interactions are positive (=repulsive), 
as shown in Fig. \ref{fig:fig4} (c).
However, as shown in Fig. \ref{fig:fig4} (d),
the intra-pocket interaction becomes negative (=attractive)
for $V^{\Lambda}(\k,\p)$ in the presence of $U$-VCs,
since the pairing force due to the orbital fluctuations 
is multiplied by $|\Lambda^c|^2\gg 1$.
Since the averaged inter-pocket interaction is tiny,
$s$-wave state and $d$-wave state are approximately degenerate.
(see Fig. \ref{fig:AP3} (b) without SOI.)
In Figs. \ref{fig:fig4} (c) and (d),
the frequency-independent (${\hat U}^0$)-linear term is dropped,
although it is included in solving the gap equation below.

Up to now, we analyzed only the single-fluctuation-exchange processes.
We also find that large attractive interaction is given by the
``AL-type crossing-fluctuation-exchange process'' ${\hat V}^{\rm cross}(k,p)$
shown in Fig. \ref{fig:fig4} (e).
Its significance is naturally expected
since ${\hat V}^{\rm cross}$ is mathematically equivalent to the AL-VC
that plays significant role in the present multiorbital system 
\cite{Onari-SCVC}.
Physically, $V^{\rm cross}(k,p)$ represents the 
pairing glue due to the ``multi-boson-exchange processes''.
In the orbital basis, its analytic expression is
\begin{eqnarray}
	&&V^{\rm cross}_{l,l',m,m'} ( k, p )
	= \frac{T}{4} \sum_{q} \sum_{a,b,c,d}
	G_{a,b} (p-q)
	G_{c,d} (-k-q)
\nonumber \\
	&& \times
	\left\{
        3 I'^{s}_{l,a;m,d} (k-p+q) I'^{s}_{b,l',c,m'} (-q) 
        \right.
\nonumber \\
	&&\quad 
	+ 3 I'^{s}_{l,a;m,d} (k-p+q) I'^{c}_{b,l',c,m'} (-q) 
\nonumber \\
	&&\quad  
        + 3 I'^{c}_{l,a;m,d} (k-p+q) I'^{s}_{b,l',c,m'} (-q) 
\nonumber \\
	&&\quad \left.
	-   I'^{c}_{l,a;m,d} (k-p+q) I'^{c}_{b,l',c,m'} (-q) 
	\right\} 
\label{eqn:Vcross} ,
\end{eqnarray}
where we put ${\hat I}'^x= {\hat I}^x-{\hat U}^{0x}$
to avoid the double counting of diagrams included in other terms.
This analytic expression is essentially the same as
that for the AL-type $U$-VC in Fig. \ref{fig:Lambda-x}.
so $V^{\rm cross}(k,p)$ is naturally expected to be important 
in Fe-based superconductors.

To understand why ${\hat V}^{\rm cross}(k,p)$ becomes negative,
we consider the case that $V^{s,c}$ has moderate energy dependence.
By dropping the orbital indices of $V^{s,c}$ for simplicity, we obtain
\begin{eqnarray}
&&V^{\rm cross}(\k,\p) \approx -\sum_{\q}
\frac{f_{\k-\q}-f_{\p-\q}}{\e_{\p-\q}-\e_{\k-\q}}
\{ 3I^s(\q)I^s(\k+\p-\q)
\nonumber \\
&&\ \ \ +3I^s(\q)I^c(\k+\p-\q)
+3I^c(\q)I^s(\k+\p-\q)
\nonumber \\
&&\ \ \ -I^c(\q)I^c(\k+\p-\q) \},
\label{eqn:Vcross-ap}
\end{eqnarray}
where $f_\k$ is the Fermi distribution function for $\e=\e_\k$.
Then, $\frac{f_{\k-\q}-f_{\p-\q}}{\e_{\p-\q}-\e_{\k-\q}}$ is always positive.
In a single-orbital model,
$I^s(\q)\approx U/(1-U\chi^{(0)})>0$ and 
$I^c(\q)\approx -U/(1+U\chi^{(0)})<0$,
so $V^{\rm cross}$ is small due to the cancellation,
which can be verified numerically 
in the single-orbital Hubbard model for cuprate superconductors.
In contrast, in the FeSe model, 
$I^c(\q)_{m,m;m,m}$ has positive value even in the RPA.
When $\chi^{\rm orb}\gg1$ due to the AL-VC,
$I^c_{2,2;2,2}(\q) \sim -U+U^2\chi^{\rm orb}(\q)/4$ 
takes large positive value for $J/U\ll1$.
Thus, $V^{\rm cross}$ gives the large negative interaction
in the multiorbital model.

According to Eq. (\ref{eqn:Vcross-ap}),
$V^{\rm cross}_{m,m;m,m}$ can take large negative value when 
the spin fluctuations develop on the $m$-orbital.
In the present FeSe model,
the magnitudes of $\chi^s_{m,m;m,m}$ for $m=2\sim4$ are comparable.
Since the $d_{xy}$-orbital is involved in both electron-pockets,
$V^{\rm cross}_{4,4;4,4}$ gives large attractive inter-pocket interaction
in the present study.

Since Eq. (\ref{eqn:Vcross-ap}) gives an over-estimated value,
we perform the serious numerical analysis based on Eq. (\ref{eqn:Vcross-ap}):
Figure \ref{fig:fig4} (f) shows the obtained ${\hat V}^{\rm cross}(\k,\p)$
on the FSs.
Since the total pairing interaction
\begin{eqnarray}
{\hat V}^{\rm tot}(k,p)= {\hat V}^{\Lambda}(k,p)+{\hat V}^{\rm cross}(k,p)
\label{eqn:Vtot} ,
\end{eqnarray}
is negative for both inter- and intra-pocket part,
the $s$-wave state should appear.
The significant role of $V^{\rm cross}$
is one of the main findings in the present study.
It is verified that the momentum-dependence of 
${\hat V}^{\rm cross}(\k,\p)$ given by Eq. (\ref{eqn:Vcross-ap})
is similar to that given by Eq. (\ref{eqn:Vcross}).
However, the former is over-estimated by two- or three-times.

\section{$S_{++}$-wave gap function in 15\% e-doped FeSe}

\begin{figure}[!htb]
\includegraphics[width=.99\linewidth]{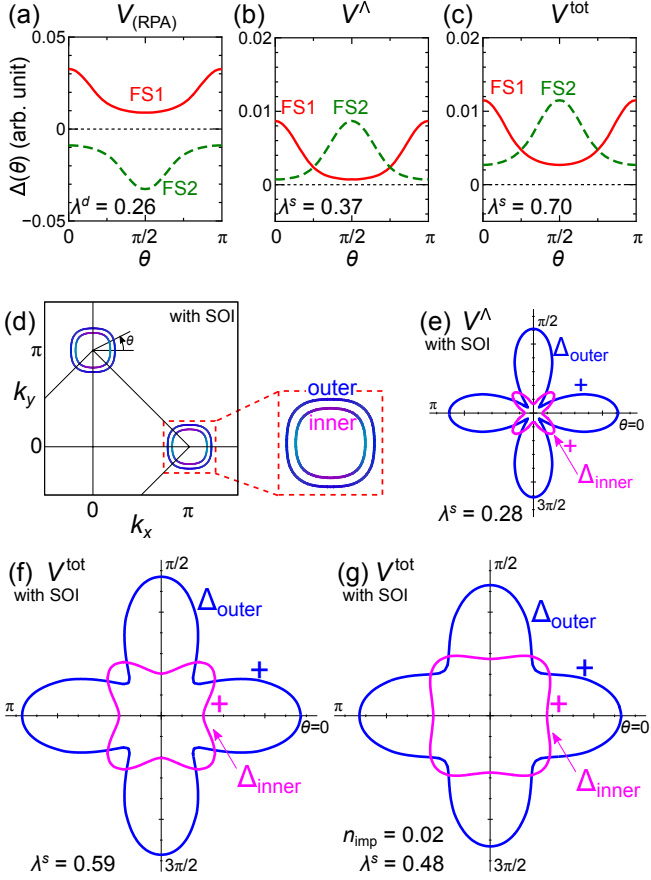}
\caption{
(color online)
The gap functions in the 1Fe-UC FeSe model obtained for the pairing interaction
(a) $V_{\rm (RPA)}$ 
(b) $V^\Lambda$ (with $U$-VC)
and (c) $V^{\rm tot}$ (with $U$-VC and crossing term),
without the SOI.
In (b) and (c), the $s$-wave state is realized 
due to the orbital fluctuations.
In all cases,  the gap function has maximum 
on the region with strong $d_{xy}$-orbital character.
(d) FSs in the 2Fe-UC FeSe model for $\lambda_{\rm SOI}=80$ meV.
The outer (inner) FS is mainly composed of the $xy$-orbital ($xz,yz$-orbitals).
We present the obtained $s_{++}$-wave gap functions for
(e) $V^\Lambda$ and (f) $V^{\rm tot}$.
(g) The $s_{++}$-wave state in the presence of impurities,
by which the gap anisotropy is smeared.
The anisotropic $s_{++}$-wave states in (f) and (g) are consistent with 
the experimental reports 
\cite{Shen-eFeSe,Feng-eFeSe-swave,Feng-eFeSe-swave2,HHWen-eFeSe}.
}
\label{fig:fig5}
\end{figure}

From now on, we analyze the gap equation (\ref{eqn:gapeq}) numerically.
We solve the frequency dependence of the gap function seriously, 
by restricting the momentum $\k,\p$ on the FSs
as done in Ref. \cite{Saito-SOI}.
%
In the absence of the SOI, 
the pairing interaction in the band-diagonal basis is
\begin{eqnarray}
V^{\rm tot}_{\a,\b}(k,p)&=& \sum_{ll'mm'} V^{\rm tot}_{l,l';m,m'}(k,p)
\nonumber \\
&&\times u^*_{l\a}(\k)u_{l'\b}(\p)u_{m\b}(-\p)u^*_{m'\a}(-\k) ,
\end{eqnarray}
where $u_{l\a}(\k)=\langle\k;l|\k;\a\rangle$ is the unitary matrix 
connecting between the band representation and the orbital one.
Then, the gap equation is rewritten as
\begin{eqnarray}
\lambda z^{-1}_\a(\k)\Delta_\a(\k,\e_n)
&=& -\frac{\pi T}{(2\pi)^2} \sum_{\b,m}\int_{\rm FS\b}\frac{d \p}{|v_{\p}^\b|}
V_{\a,\b}^{\rm tot}(\k,\e_n,\p,\e_m)
\nonumber \\
& &\times \frac{\Delta_\b(\p,\e_m)}{|\e_m|}
\label{eqn:gapeq2} ,
\end{eqnarray}
where $\lambda$ is the eigenvalue.
$z_\a(\k)=\sum_l z_l|u_{l,\a}(\k)|^2$
is the renormalization factor for band $\a$.
The gap equation in the presence of the SOI is explained 
in Sect. II C of Ref. \cite{Saito-SOI}.
In the numerical study,
we calculate $U$-VCs in $V^{\rm tot}$ only for $|\e_n|=|\e_m|=\pi T$,
and put ${\hat \Lambda}^{c,s}={\hat 1}$ for others.
This simplification is unfavorable for obtaining the 
$s_{++}$-wave state.
Nonetheless of this underestimation,
the $s_{++}$-wave state is realized in Fig. \ref{fig:fig5} (f)
in the main text.

First, we study the 1Fe-UC model without the SOI
shown in Fig. \ref{fig:fig2} (a):
In the RPA without any $U$-VC, $V^{\rm SC}=V_{\rm (RPA)}$,
the spin-fluctuation-mediated $d$-wave state is obtained 
in Fig. \ref{fig:fig5} (a).
Here, $|\Delta(\theta)|$ takes maximum on the region with 
large $d_{xy}$-orbital weight due to large
spin fluctuations on the $d_{xy}$-orbital.
However, the eigenvalue for the $d$-wave 
is just $\lambda^d=0.26$ since the spin fluctuations are weak.

On the other hand, the $s$-wave state is obtained if 
the $U$-VC is taken into account, $V^{\rm SC}=V^{\Lambda}$,
shown in Fig. \ref{fig:fig5} (b).
Here, the pairing interaction in Eq. (\ref{eqn:V-UVC}),
which is shown in Fig. \ref{fig:fig4} (d), is
given by $\Lambda^{s,c}$ and ${\hat \chi}^{s,c}(q)$ obtained by the SC-VC theory.
However, eigenvalue is still small ($\lambda^s=0.37$)
due to the cancellation in the inter-pocket interaction;
see Figs. \ref{fig:fig4} (d).
As we show in Fig. \ref{fig:fig5} (c),
the $s$-wave state with large eigenvalue ($\lambda^s=0.70$) 
is obtained for the total pairing interaction $V^{\rm tot}=V^\Lambda+V^{\rm cross}$ 
in Eq. (\ref{eqn:Vtot}),
because of the attractive force by the crossing term $V^{\rm cross}$.
Here, $|\Delta(\theta)|$ takes maximum on the
$d_{xy}$-orbital character region
because of the repulsive intra-pocket interaction on the $d_{xz(yz)}$-orbital
due to $\chi^s_{xz(yz)}(\q)$ with small-$q$,
in addition to the attractive inter-pocket interaction 
due to $V^{\rm cross}(k,p)$ on the $d_{xy}$-orbital.

In the next stage, 
we analyze the gap equation in the 2Fe-UC FeSe model with the SOI,
by following the theoretical procedure in 
Ref. \cite{Saito-SOI}.
Figure \ref{fig:fig5} (d) shows the FSs for $\lambda_{\rm SOI}=80$ meV.
Due to the SOI-induced hybridization,
very anisotropic $s_{++}$-wave state is obtained for $V^{\rm SC}=V^{\Lambda}$, 
as shown in Fig. \ref{fig:fig5} (e).
Figure \ref{fig:fig5} (f) shows the moderately anisotropic 
$s_{++}$-wave state obtained for $V^{\rm SC}=V^{\rm tot}$ [$\lambda^s=0.59$].
The anisotropy of the $s_{++}$-wave state is smeared out 
by introducing small amount of the impurity as well-known:
Figure \ref{fig:fig5} (g) show the $s_{++}$-wave state
in the presence of the 2\% impurity with the constant inter- and intra-pocket
scattering potential $I_{\rm inter}=I_{\rm intra}=1$ eV in the Born approximation.
Thus, $s_{++}$-wave state with large $\lambda^s$
is obtained under moderate spin and orbital fluctuations.
The anisotropic $s$-wave states in Figs. \ref{fig:fig5} (f) and (g)
are consistent with the experimental reports in 
Refs. \cite{Shen-eFeSe,Feng-eFeSe-swave,Feng-eFeSe-swave2,HHWen-eFeSe}.

\section{Robustness of the $s$-wave state 
}

In the previous section,
we performed the numerical study of the gap equation 
for the 15\% $e$-doped FeSe model.
The obtained anisotropic $s_{++}$-wave state shown in Fig. \ref{fig:fig5}
is quantitatively consistent with experiments.
In Fig. \ref{fig:fig5},
we showed the numerical results 
only for $r=0.35$ at $T=30$ meV.

\begin{figure}[!htb]
\includegraphics[width=.99\linewidth]{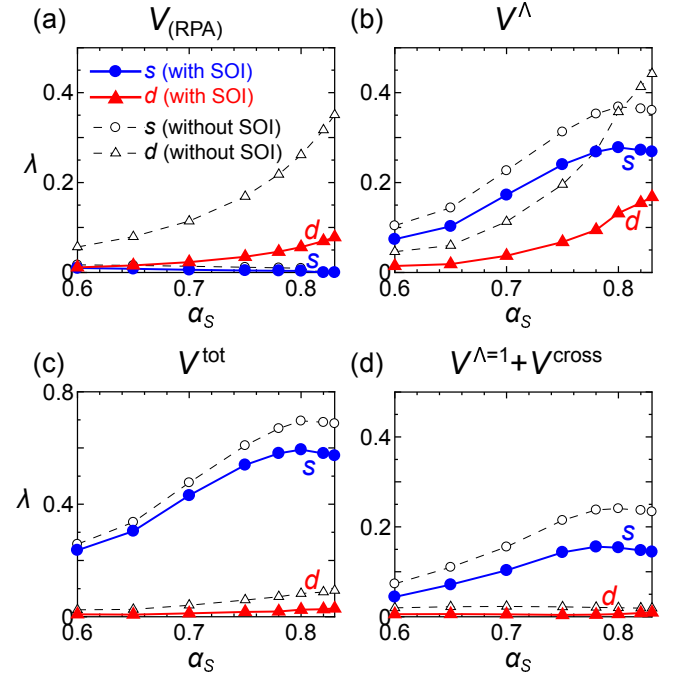}
\caption{
(color online)
The eigenvalue for the $s$-wave state $\lambda^s$ 
and that for the $d$-wave state $\lambda^d$ obtained for the 
pairing interaction (a) $V_{\rm (RPA)}$,
(b) $V^{\Lambda}$, and (c) $V^{\rm tot}$.
We also show the eigenvalues
for (d) $V^{\rm SC} =V^{\Lambda=1}+V^{\rm cross}$.
}
\label{fig:AP3}
\end{figure}

Here, we explain that the $s_{++}$-wave state
is obtained for wide parameter range.
Figure \ref{fig:AP3} 
shows eigenvalues for the pairing interaction 
(a) $V_{\rm (RPA)}$, (b) $V^{\Lambda}$, (c) $V^{\rm tot}$,
and (d) $V^{\Lambda=1}+V^{\rm cross}$
as functions of $\a_S$.
In the RPA without any $U$-VC in Fig. \ref{fig:AP3} (a),
the $d$-wave state is realized stably,
whereas the eigenvalue for the $d$-wave state, $\lambda^d$,
is strongly suppressed by the SOI.
In Fig. \ref{fig:AP3} (b), we taking the $U$-VC into account.
Due to the $U$-VC,
$\lambda^s$ increases drastically whereas $\lambda^d$ is qualitatively unchanged
compared to Fig. \ref{fig:AP3} (a).
The reason is that the attractive 
intra-pocket force is enlarged by $|\Lambda^c|^2\gg1$, whereas
the repulsive inter-pocket force is suppressed by $|\Lambda^s|^2\ll1$.
In the presence of the SOI,
the $d$-wave state is drastically suppressed, 
so the relation $\lambda^s>\lambda^d$ is realized.

The crossing term $V^{\rm cross}$ in Fig. \ref{fig:fig4} (e)
or Eq. (\ref{eqn:Vcross})
gives large inter-pocket attractive interaction.
Figure \ref{fig:AP3} (c) shows the eigenvalues for 
the pairing interaction $V^{\rm tot}=V^{\Lambda}+V^{\rm cross}$. 
Thanks to the strong attractive interaction by $V^{\rm cross}$,
$\lambda^s$ is largely enlarged even in the absence of the SOI.
Thus, the $s_{++}$-wave state is realized for wide range of $\a_S$
by going beyond the ME approximation.

We also show the $s$-wave and $d$-wave eigenvalues 
for $V^{\rm SC} =V^{\Lambda=1}+V^{\rm cross}$ in Fig. \ref{fig:AP3} (d).
($V^{\Lambda=1}$ represents the Migdal approximation.
For $V^{\rm cross}$,
we replace each ${\hat I}'$ in Eq. (\ref{eqn:Vcross}) with ${\hat I}$,
and subtract the $({\hat U}^0)^2$-term.)
In this case, the $s$-wave state is realized 
because of the attractive interaction by $V^{\rm cross}$.
However, the obtained $\lambda^s$ is just $\sim0.3$.
Thus, both $U$-VC and $V^{\rm cross}$ are necessary
for explaining the fully-gapped $s$-wave state with large $\lambda^s$.

In summary, 
the $s_{++}$-wave state is obtained for wide parameter range
in Fig. \ref{fig:AP3} (c)
due to the combination of the $U$-VC and crossing term,
even if the SOI is neglected.
Once the SOI is taken into account correctly,
the $s_{++}$-wave state is realized even if 
we drop either $U$-VC or $V^{\rm cross}$ but not both.

\section{Discussions}

The mechanism of high-$T_{\rm c}$ superconductivity 
in Fe-based superconductors still remains an open problem.
To attack this important issue,
the high-$T_{\rm c}$ state ($T_{\rm c}\gtrsim60$K) without hole-pockets
in heavily $e$-doped FeSe provides an excellent opportunity,
since its very simple bandstructure 
is favorable for the unambiguous theoretical study.
The only possible pairing states are $s$-wave state and $d$-wave state, and
high-$T_{\rm c}$ state is realized irrespective of the small 
spin fluctuations \cite{Koike-eFeSe}.
Similar high-$T_{\rm c}$ with small spin fluctuations
is realized in F- and H-doped LaFeAsO \cite{Fujiwara}.

We analyzed the realistic Hubbard model for FeSe using the SC-VC theory,
and found that moderately developed spin and orbital fluctuations appear
for 15\% $e$-doped case,
consistently with the experimental phase diagram
\cite{Feng-eFeSe}.
Next, we uncovered two significant ``beyond-ME processes''
for high-$T_{\rm c}$ pairing mechanism:
(i) VC for the electron-boson coupling ($U$-VC), and
(ii) AL-type crossing-fluctuation-exchange term ($V^{\rm cross}$).
%
Due to (i), strong intra-pocket attractive interaction is caused 
even when the ferro-orbital fluctuations are moderate.
For this reason, $T_{\rm c}$ is enlarged for both $d$-wave and $s$-wave states.
We stress that the phonon-mediated attractive interaction is also 
enlarged by the $U$-VC \cite{Tazai}, so it is important to study
the increment of $T_{\rm c}$ due to the electron-phonon mechanism
\cite{Shen-replica,Millis,DHLee}.
Due to (ii), large inter-electron-pocket attractive interaction is realized.
$V^{\rm cross}(k,p)$ represents the pairing interaction due to the 
``multi-fluctuation-exchange processes''.
Due to these beyond-ME processes,
the fully-gapped $s_{++}$-wave state is satisfactorily explained.
The obtained anisotropic gap structure is 
consistent with experimental results.

The significance of the $U$-VC in $V^\Lambda$ 
has been verified in various multiorbital Hubbard models 
\cite{Onari-SCVCS,Yamakawa-bulkFeSe,Tazai}.
In Ref. \cite{Tazai}, we verified the 
significance of the $U$-VC by applying 
the functional-renormalization-group (fRG) method 
to the two-orbital Hubbard model:
We showed that both the ``$\k$-dependence'' and the 
``spin/charge-channel dependence'' 
of the pairing interaction given by the fRG four-point vertex
are well approximated by the ``single-fluctuation-exchange approximation 
with the $U$-VCs''. 
Although $V^{\rm cross}$ gives large attractive interaction in 
electron-doped FeSe model, its importance seems to be model-dependent. 
In fact, $V^{\rm cross}$ is less important in the two-orbital model \cite{Tazai} 
and in the undoped FeSe model studied in Ref. \cite{Yamakawa-bulkFeSe}.
It is our important future problem to clarify
the importance of the multi-fluctuation-exchange interactions
in FeSe and other models, by using the fRG theory.

In Appendix C,
we analyze the bulk FeSe model ${\hat H}^{0,{\rm b}}_\k$ ($\Delta H^0=0$)
with 15\% $e$-doping, and obtain 
the full-gap $s_{++}$-wave state similar to Fig. \ref{fig:fig5} (f).
However, the eigenvalue is only $\lambda^s=0.22$.
This result indicates that the change in the bandstructure
in monolayer FeSe observed by ARPES, which is realized by 
lifting the $d_{xy}$-orbital level at X,Y points
shown in Fig. \ref{fig:fig2} (a),
is important to realize high-$T_{\rm c}$ superconductivity.

The present gap equation beyond the standard ME formalism 
should be useful for understanding the 
rich variety of the superconducting states in Fe-based superconductors.
The proposed ``inter-electron-pocket pairing mechanism''
will enlarge $T_{\rm c}$ in other Fe-based superconductors,
even if the $s_\pm$-wave state is realized.

\acknowledgments
We are grateful to A. Chubukov, P.J. Hirschfeld, R. Fernandes,
J. Schmalian, Y. Matsuda, S. Shibauchi, and S. Onari 
for useful discussions.
This study has been supported by Grants-in-Aid for Scientific 
Research from MEXT of Japan.

\appendix

\section{Analytic expressions for the $U$-VC in the SC-VC Theory
\label{sec:B}}

Here, we present the analytic expressions for the
charge- and spin-channel $U$-VCs in the SC-VC theory.
In the SC-VC theory, the $U$-VC for $x$-channel ($x=s,c$) is given as
\begin{eqnarray}
{\hat \Lambda}^x(k,k')= {\hat 1}+{\hat \Lambda}^{{\rm MT},x}(k,k')
+{\hat \Lambda}^{{\rm AL},x}(k,k'),
\end{eqnarray}
where the index MT (AL) represents the MT (AL) term,
shown in Fig. \ref{fig:Lambda-x} in the main text.

First, we explain the MT-type $U$-VCs,
which was already introduced in Ref. \cite{Tazai}.
The charge- and spin-channel MT-terms are given as
\begin{eqnarray}
	\Lambda^{{\rm MT}, c}_{l,l';m,m'} (k,k')
  	&=& \frac{T}{2} \sum_{p} \sum_{a,b}
	\left\{
		I^{c}_{b,l';a,l} (p) + 3 I^{s}_{b,l';a,l} (p)
	\right\}
\nonumber \\
	&&\times G_{a,m} (k+p) G_{m',b} (k'+p) 
\label{eqn:UMTc} ,
\end{eqnarray}

\begin{eqnarray}
	\Lambda^{{\rm MT}, s}_{l,l';m,m'} (k,k')
  	&=& \frac{T}{2} \sum_{p} \sum_{a,b}
	\left\{
		I^{c}_{b,l';a,l} (p) - I^{s}_{b,l';a,l} (p)
	\right\}
\nonumber \\
	&&\times G_{a,m} (k+p) G_{m',b} (k'+p) 
\label{eqn:UMTs} ,
\end{eqnarray}
where
${\hat I}^x(q)= {\hat U}^{0x}{\hat \chi}^x(q){\hat U}^{0x}+{\hat U}^{0x}$,
and $a,b,l,l',m,m'$ are orbital indices.

Next, we explain the AL-type $U$-VCs,
which was also introduced in Ref. \cite{Tazai}:
The charge- and spin-channel AL-terms are given as
\begin{eqnarray}
	&&\Lambda^{{\rm AL}, c}_{l,l';m,m'} (k,k')
\nonumber \\
&&\quad = \frac{T}{2} \sum_{p} \sum_{a,b,c,d,e,f}
	G_{a,b} (k'-p) {\Lambda^0}'_{m,m';c,d;e,f} (k-k',p)
\nonumber \\
	&&\quad \times
	\left\{
	    I^{c}_{l,a;c,d} (k-k'+p) I^{c}_{b,l';e,f} (-p) \right.
\nonumber \\
	&&\quad \left. + 3 I^{s}_{l,a;c,d} (k-k'+p) I^{s}_{b,l';e,f} (-p)
	\right\}
\label{eqn:UALc} ,
\end{eqnarray}
\begin{eqnarray}
	&&\Lambda^{{\rm AL}, s}_{l,l';m,m'} (k,k')
\nonumber \\
	&&\quad = \frac{T}{2} \sum_{p} \sum_{a,b,c,d,e,f} G_{a,b} (k'-p) 
	{\Lambda^0}'_{m,m';c,d;e,f} (k-k',p)
\nonumber \\
	&& \quad \times
	\left\{
	  I^{c}_{l,a;c,d} (k-k'+p) I^{s}_{b,l';e,f} (-p)
	\right.	
\nonumber \\
	&& \quad \left.
	+ I^{s}_{l,a;c,d} (k-k'+p) I^{c}_{b,l';e,f} (-p)
	\right\}
\nonumber \\
	&& \quad
+ \delta\Lambda^{{\rm AL}, s}_{l,l';m,m'} (k,k')
%
\label{eqn:UALs} ,
\end{eqnarray}
where 
the three-point vertex ${\hat \Lambda}^0(q,p)$ is given as
\begin{eqnarray}
&&\Lambda_{l,l';a,b;e,f}^0(q,p)
\nonumber \\
&&\ \ \ =-T\sum_{k'}G_{l,a}(k'+q)G_{f,l'}(k')G_{b,e}(k'-p),
\end{eqnarray}
%
and 
${\Lambda^0}_{m,m';c,d;g,h}'(q,p)\equiv
\Lambda^0_{c,h;m,g;d,m'}(q,p)+\Lambda^0_{g,d;m,c;h,m'}(q,-p-q)$.
The last term in Eq. (\ref{eqn:UALs}) is given as
\begin{eqnarray}
	&&\delta \Lambda^{{\rm AL}, s}_{l,l';m,m'} (k,k')
	= T \sum_{p} \sum_{a,b,c,d,e,f} G_{a,b} (k'-p) 
 \nonumber \\
	&&\quad \times I^{s}_{l,a;c,d} (k-k'+p) I^{s}_{b,l';e,f} (-p)
	{\Lambda^0}''_{m,m';c,d;e,f} (k-k',p) ,
\nonumber \\
\label{eqn:del-Lambda-ALs} 
\end{eqnarray}
a
where
${\Lambda^0}''_{m,m';c,d;g,h}(q,p)\equiv
\Lambda^0_{c,h;m,g;d,m'}(q,p)-\Lambda^0_{g,d;m,c;h,m'}(q,-p-q)$.
We verified that the contribution from
Eq. (\ref{eqn:del-Lambda-ALs}) is very small.

\begin{figure}[!htb]
\includegraphics[width=.9\linewidth]{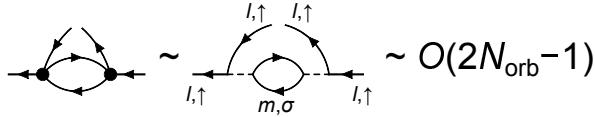}
\caption{
(color online)
The $(U^0)^2$-term for the $U$-VC, 
by which the spin-channel $U$-VC is suppressed.
One bubble scales as $\sim O(2N_{\rm orb}-1)$,
where $N_{\rm orb}$ is the $d$-orbital degrees of freedom.
}
\label{fig:AP2}
\end{figure}

Figure \ref{fig:Lambda-x} represents the 
simplified diagrammatic expression for the $U$-VC,
which is irreducible with respect to ${\hat U}^{0s(c)}$.
The ($U^0$)-linear terms in Eqs. (\ref{eqn:UMTc}) and (\ref{eqn:UMTs})
should be dropped to avoid the double counting of the RPA-type diagrams.
We also carefully drop the double counting $(U^0)^2$-terms
included in both MT and AL terms.

We verified numerically that the large charge-channel $U$-VC
($|\Lambda^c|^2\gg1$) originates from the $\chi^s$-square term 
in Eq. (\ref{eqn:UALc}).
We also verified that the 
relation $|\Lambda^s|^2\ll1$ reported in the main text 
is mainly given by the $(U^0)^2$-term in Fig. \ref{fig:AP2},
since one bubble scales as $\sim O(2N_{\rm orb}-1)$.
This result is consistent with the previous analysis for the two-orbital model
\cite{Tazai}.

We stress that the $U$-VCs is important only for low-frequencies
$\w\lesssim {\rm min}\{\w_{\rm sf},\w_{\rm cf}\}$, where $\w_{\rm sf(cf)}$ 
is the characteristic spin (orbital/charge) fluctuation energy.
Note that $\w_{\rm sf(cf)}$ becomes smaller near the 
quantum-critical-point in proportion to $1-\a_{S(C)}$. 
Since the Cooper pair is formed by low-energy electrons,
the $U$-VCs in ${\hat V}^\Lambda(k,p)$ play significant role on the 
superconducting state.
On the other hand, the 
contribution of the $U$-VCs for the two susceptibilities 
in $V^{\rm cross}$ is expected to be small because of the
frequency summation inside of $V^{\rm cross}$:
We verified this fact numerically in the present model.

\section{Susceptibilities in the SC-VC theory
}

In the main text,
we studied the FeSe model by applying the SC-VC theory
introduced in Refs. \cite{Onari-SCVC,Yamakawa-FeSe}.
We analyzed the MT-VC and AL-VC for both spin- and charge-channels
self-consistently.
In the present theory, the charge (spin) susceptibilities are given as
\begin{eqnarray}
{\hat \chi}^{c(s)}(\q)= 
{\hat \Phi}^{c(s)}(\q)({\hat 1}-{\hat \Gamma}^{c(s)}{\hat \Phi}^{c(s)}(\q))^{-1}
\label{eqn:chisc2}
\end{eqnarray}
where ${\hat \Phi}^{c(s)}(\q)$ is given in Eq. (\ref{eqn:Phi}) in the main text.
It is rewritten as
${\hat \Phi}^{x}(\q)={\hat \chi}^{0}(\q)+
{\hat X}^{{\rm MT},x}(\q)+{\hat X}^{{\rm AL},x}(\q)$,
where $X^{{\rm MT(AL)},x}$ represents the MT-VC (AL-VC).

The charge- and spin-channel AL-VCs are given as
\begin{eqnarray}
&&X_{l,l';m,m'}^{{\rm AL},c}(q)=\frac{T}2\sum_{p}\sum_{a\sim h}
\Lambda^0_{l,l';a,b;e,f}(q,p)
\nonumber \\
&&\!\!\!\!\!\!\!\!\!
 \ \ \times \{  3{I}_{a,b;c,d}^s(p+q){I}_{e,f;g,h}^s(-p)
+{I}_{a,b;c,d}^c(p+q){I}_{e,f;g,h}^c(-p) \}
\nonumber \\
&&\!\!\!\!\!\!\!\!\!
 \ \ \times {\Lambda^0}_{m,m';c,d;g,h}'(q,p) ,
\label{eqn:ALc} \\
&&X_{l,l';m,m'}^{{\rm AL},s}(q)=\frac{T}{2}\sum_p\sum_{a\sim h}\Lambda^0_{l,l';a,b;e,f}(q,p)
\nonumber \\
&& \ \ \ \times \{I^c_{a,b;c,d}(p+q)I^s_{e,f;g,h}(-p)\nonumber\\
&& \ \ \ \ +I^s_{a,b;c,d}(p+q)I^c_{e,f;g,h}(-p)\}
{\Lambda^0}'_{m,m';c,d;g,h}(q,p)
\nonumber \\
&& \ \ \ \ + \delta X_{l,l';m,m'}^{{\rm AL},s}(q)
\label{eqn:ALs} ,
\end{eqnarray}
where
${\hat I}^x(q)= {\hat U}^{0x}{\hat \chi}^x(q){\hat U}^{0x}+{\hat U}^{0x}$,
and $a\sim f$ are orbital indices.
The three-point vertex $\Lambda_{l,l';a,b;e,f}^0(q,p)$ is given as
$-T\sum_{k'}G_{l,a}(k'+q)G_{f,l'}(k')G_{b,e}(k'-p)$.
Also, ${\Lambda^0}_{m,m';c,d;g,h}'(q,p)\equiv
\Lambda^0_{c,h;m,g;d,m'}(q,p)+\Lambda^0_{g,d;m,c;h,m'}(q,-p-q)$.
The last term in Eq. (\ref{eqn:ALs}) is given as
\begin{eqnarray}
&&\delta X_{l,l';m,m'}^{{\rm AL},s}(q)=
T \sum_p\sum_{a\sim h}\Lambda^0_{l,l';a,b;e,f}(q,p)
\nonumber \\
&& \ \ \ \times I^s_{a,b;c,d}(p+q)I^s_{e,f;g,h}(-p){\Lambda^0}''_{m,m';c,d;g,h}(q,p),
\end{eqnarray}
where ${\Lambda^0}''_{m,m';c,d;g,h}(q,p)\equiv
\Lambda^0_{c,h;m,g;d,m'}(q,p)-\Lambda^0_{g,d;m,c;h,m'}(q,-p-q)$.
This term is found to be very small.

The expressions of the charge- and spin-channel MT-VCs 
are given in Ref. \cite{Text-SCVC}.
The double-counting second-order terms with respect to $H_U$
in ${\hat X}^{{\rm MT},s(c)}+{\hat X}^{{\rm AL},s(c)}$
should be subtracted to obtain reliable results \cite{Text-SCVC}.

\begin{figure}[!htb]
\includegraphics[width=.7\linewidth]{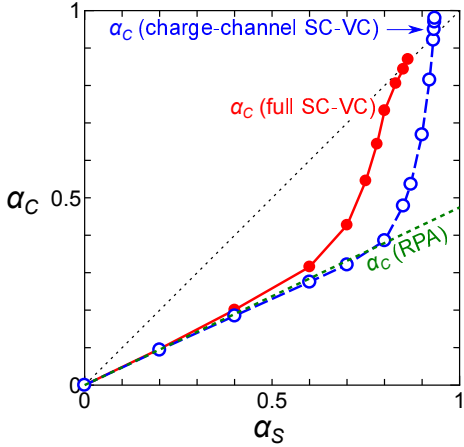}
\caption{
(color online)
Stoner factors obtained by the ``full SC-VC'' method
performed in the main text.
For comparison, the result given by the ``charge-channel SC-VC''
and that by the RPA are also shown.
}
\label{fig:AP1}
\end{figure}

Figure \ref{fig:AP1} shows the relation between the 
spin and charge Stoner factors, $\a_S$ and $\a_C$,
in the 15\% $e$-doped FeSe model
obtained by the SC-VC theory for $r=0 \sim 0.35$.
Both spin- and charge-channel VCs are calculated self-consistently.
Both $\a_S$ and $\a_C$ increase with $r$ monotonically,
and $\a_C$ exceeds $\a_S$ for $\a_{S,C}=0.86$,
since $\chi^s$-square term in the charge-channel AL-VC 
in Eq. (\ref{eqn:ALc}) becomes significant for $\a_S\rightarrow1$. 
For comparison, we show the Stoner factors
in the case that the spin-channel VC is dropped and
only the charge-channel VC are studied self-consistently,
shown as the ``charge-channel SC-VC'' method.
The obtained result is essentially similar to the 
``full SC-VC'' method performed in the main text.

\section{Analysis of the original bulk FeSe tight-binding model
}

In the main text, 
we analyzed the heavily $e$-doped FeSe model
introduced in the Method section.
To reproduce the experimental bandstructure for high-$T_{\rm c}$ FeSe
with heavily $e$-doping, we introduced the additional term $\Delta H_0$
into the tight-binding model for bulk FeSe,
as in Eq. (\ref{eqn:H0}).
Here, we perform the same analysis for 
the original bulk FeSe tight-binding model ($\Delta H_0=0$)
with 15\% $e$-doping, and obtain the $s_{++}$-wave state 
that is similar to Fig. 5 in the main text.

\begin{figure}[!htb]
\includegraphics[width=.99\linewidth]{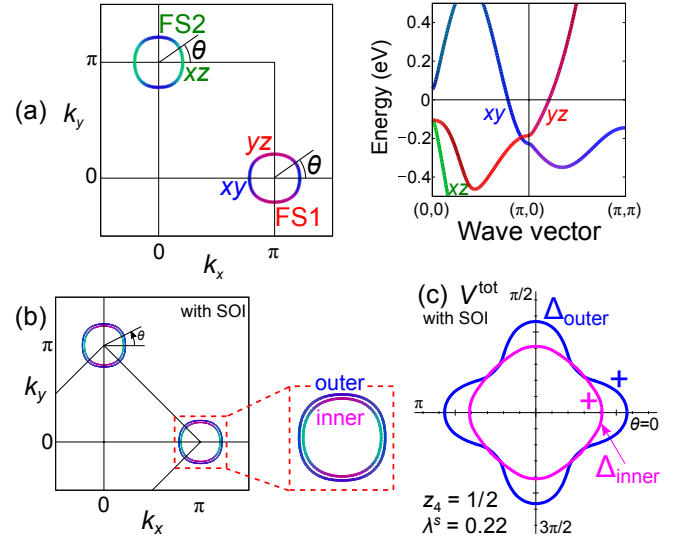}
\caption{
(color online)
(a) FSs and bandstructure of the bulk FeSe model (${\hat H}^{0,{\rm b}}_\k$)
with 15\% $e$-doping.
Green, red, and blue lines correspond to $xz$, $yz$, and $xy$ orbitals, 
respectively.
(b) FSs in the 2Fe-UC FeSe model for $\lambda_{\rm SOI}=80$ meV.
The outer (inner) FS is mainly composed of the $xy$-orbital ($xz,yz$-orbitals).
(c) The obtained $s_{++}$-wave gap function for $V^{\rm tot}$
in the case of $r=0.36$ and $z_4=1/2$; $(\a_S,\a_C)=(0.83,0.85)$.
}
\label{fig:AP4}
\end{figure}

Figure \ref{fig:AP4} (a) shows the FSs and bandstructure
for the bulk FeSe model ($\Delta H_0=0$) with 15\% $e$-doping.
Here, the relation $E_{xy}<E_{yz}$ holds at X-point.
The shape of the FSs is similar to that for the 
heavily $e$-doped FeSe model in the main text, shown in Fig. 2 (a).
On the other hand, the Fermi velocity in Fig. \ref{fig:AP4} (a) 
is larger compared to Fig. 2 (a) in the main text.
For this reason, the density-of-states at the Fermi level
is relatively small in the present  bulk FeSe tight-binding model.
By applying the SC-VC method, 
we obtain the moderate incommensurate spin fluctuations
and ferro-orbital fluctuations,
which are essentially similar to thoese in the main text 
shown in Fig. 2(c).
Figure \ref{fig:AP4} (b) shows the FSs 
in the 2Fe-UC model for $\lambda_{\rm SOI}=80$ meV.
The outer (inner) FS is mainly composed of the $xy$-orbital ($xz,yz$-orbitals).
Next, we study the superconducting state
for the total pairing interaction $V^{\rm tot}$
given by the SC-VC thoery.
The obtained full-gap $s_{++}$-wave state is shown in Fig. \ref{fig:AP4} (c).
Here, we put $r=0.36$ and $z_4=1/2$ at $T=30$ meV, 
in which the Stoner factors are $(\a_S,\a_C)=(0.83,0.85)$.

Therefore, the full-gap $s_{++}$-wave state similar to Fig. \ref{fig:fig5} (f) 
in the main text
is obtained by analyzing the bulk FeSe model ${\hat H}^{0,{\rm b}}_\k$
with 15\% $e$-doping.
On the other hand, the obtained eigevalue is just $\lambda^s=0.22$.
This result indicates that the change in the bandstructure
in monolayer FeSe, which is reproduced by $\Delta H_0$ in the present Hamiltonian,
is important to realize high-$T_{\rm c}$ superconductivity.


\end{document}